\documentclass[showpacs,amsmath,amssymb,pre,preprint]{revtex4}
\usepackage{graphicx}
\usepackage{bm}

\begin{document}

\title{Quantum kinetic theory of the filamentation instability}

\author{A. Bret}
\affiliation{ETSI Industriales, Universidad de Castilla-La Mancha, 13071 Ciudad Real, Spain}

\author{F. Haas}
\affiliation{Departamento de F{\'i}sica, Universidade Federal do Paran\'a, 81531-990, Curitiba, Paran\'a, Brazil}

\begin{abstract}
The quantum electromagnetic dielectric tensor for a multi species plasma is re-derived from the gauge invariant Wigner-Maxwell system and presented under a form very similar to the classical one. The resulting expression is then applied to a quantum kinetic theory of the electromagnetic filamentation instability. Comparison is made with the quantum fluid theory including a Bohm pressure term, and with the cold classical plasma result. A number of analytical expressions are derived for the cutoff wave vector, the largest growth rate and the most unstable wave vector.
\end{abstract}


\maketitle

\section{Introduction}
In the last years there has been an increasing interest on electromagnetic (as opposed to purely electrostatic) quantum plasma phenomena. With the emergence of new fields like spintronics \cite{zutic} where magnetic effects
are fundamental, it is essential to have a deeper knowledge on electromagnetic quantum plasma models. For instance, recently Eliasson and Shukla \cite{eliasson} have reported on the Bernstein modes in a magnetized, degenerated quantum plasma described by the Vlasov-Maxwell system. In this work, the fermionic nature of the charge carriers was taken into account thanks to the underlying Fermi-Dirac type equilibrium. Consequently, the upper hybrid dispersion relation for a degenerate quantum plasma was derived. However, the employed kinetic equation was classical (Vlasov), so that the second quantum effect besides statistics, or quantum diffraction effects, was not included. To take care of the typical quantum phenomena like wave packet spreading and tunneling, one must resort to a quantum kinetic equation for the reduced one-particle Wigner function, which is the quantum analog of the usual classical reduced one-particle distribution function in phase space. For these reasons, it is advisable to go one step further and formulate a general kinetic electromagnetic theory for linear waves in quantum plasmas. There are, nevertheless, already many instances where linear and nonlinear wave propagation in electromagnetic quantum plasmas were treated. For instance, one can cite the quantum Weibel instability, the dense plasma magnetization by the electromagnetic waves, the temporal dynamics of spins in magnetized plasmas, stimulated scattering quantum instabilities and the analysis of self-trapped electromagnetic waves in a quantum hole, as reviewed in Ref. \cite{uspekhi} by Shukla and Eliasson.

In the present work we show a kinetic treatment of the quantum filamentation instability, significantly generalizing the earlier fluid-based model \cite{BretPoPQuantum2007,BretPoPQuantum2008}. Moreover, the basic kinetic model we use is the evolution equation for the gauge-invariant Wigner function \cite{Stratonovich}, which fully takes into account quantum diffraction. In addition, we assume a counter-streaming zero-temperature Fermi-Dirac equilibrium, so that the quantum statistical effects due to the exclusion principle are also incorporated.

This work is organized as follows. In Section II, we derive the general electromagnetic dispersion relation for linear waves in quantum plasmas. In Section III, the quantum kinetic filamentation instability is worked out. The dispersion relation is compared with the results from the classical zero-temperature fluid model and the quantum hydrodynamic model, with a Bohm potential term. Also the behaviors of the cuttof wave vector, of the most unstable wave vector and the corresponding largest growth rate are studied. Further, analytic estimates for these quantities are provided. Section IV is reserved to the conclusions.

\section{Dielectric tensor}
The electromagnetic dielectric tensor of a collisionless classical plasma can be derived in a quite standard manner from the Vlasov-Maxwell system \cite{Ichimaru, Clemmow}. In a very similar way, the quantum version of the same tensor can be derived from the Wigner-Vlasov system, where the Wigner equation \cite{Wigner} is the quantum equivalent to the Vlasov one. Following Ref. \cite{HaasNJP2010},  the gauge invariant Wigner function $f_{j}({\bf r}, {\bf v}, t)$ for particles of species $j$ with charge $q_j$ and mass $m_j$, obeys a Vlasov-like equation introduced by Stratonovich \cite{Stratonovich}. This same equation was put into an illuminating form by Serimaa \cite{Serimaa}, according to
\begin{equation}
\label{e11}
\left(\frac{\partial}{\partial t} + ({\bf v} + \Delta\tilde{\bf v}_j)\cdot\frac{\partial}{\partial{\bf r}} + \frac{q_j}{m_j}\,\left[\tilde{\bf E}_j + ({\bf v} + \Delta\,\tilde{\bf v}_j)\times\tilde{\bf B}_j \right]\,\cdot\frac{\partial}{\partial{\bf v}}
\right)\,f_{j}({\bf r},{\bf v},t) = 0.
\end{equation}
In Eq. (\ref{e11}) we have the following differential operators,
\begin{eqnarray}
\label{e8}
\Delta\tilde{\bf v}_j &=&  \frac{i\,\hbar\,q_j}{m^{2}_j}\,\frac{\partial}{\partial{\bf v}}\times\int_{-1/2}^{1/2}d\tau\,\tau{\bf B}\left({\bf r}+\frac{i\,\hbar\,\tau}{m_j}\,\frac{\partial}{\partial{\bf v}},\,t\right) \,,\\
\label{e9}
\tilde{\bf E}_j &=& \int_{-1/2}^{1/2}d\tau\,{\bf E}\left({\bf r}+\frac{i\,\hbar\,\tau}{m_j}\,\frac{\partial}{\partial{\bf v}},\,t\right) \,,\\
\label{e10}
\tilde{\bf B}_j &=& \int_{-1/2}^{1/2}d\tau\,{\bf B}\left({\bf r}+\frac{i\,\hbar\,\tau}{m_j}\,\frac{\partial}{\partial{\bf v}},\,t\right) \,,
\end{eqnarray}
where ${\bf B} = {\bf B}({\bf r}, t)$ and ${\bf E} = {\bf E}({\bf r}, t)$ are the magnetic and electric fields respectively. To compute $\Delta\tilde{\bf v}_j, \tilde{\bf E}_j$ and $\tilde{\bf B}_j$, one has first to Taylor-expand in powers of $\hbar$ the electromagnetic fields in the integrands and then perform the integrals. For instance, up to second-order in $\hbar$, one has
\begin{eqnarray}
\tilde{E}_{j\alpha} &=& E_{j\alpha} - \frac{\hbar^2}{24m_{j}^2}\sum_{\beta,\gamma}\frac{\partial^2 E_{j\alpha}}{\partial r_\beta \partial r_\gamma}\frac{\partial^2}{\partial v_\beta\,\partial v_\gamma} + \dots \,,\\
\tilde{B}_{j\alpha} &=& B_{j\alpha} - \frac{\hbar^2}{24m_{j}^2}\sum_{\beta,\gamma}\frac{\partial^2 B_{j\alpha}}{\partial r_\beta \partial r_\gamma} \frac{\partial^2}{\partial v_\beta\,\partial v_\gamma} + \dots \,.
\end{eqnarray}
The kinetic equation (\ref{e11}) follows from the von Neumann equation solved by the reduced one-body density matrix.  As it stands, Eq. (\ref{e11}) is manifestly gauge-invariant since it depends only on the fields.

Assuming an equilibrium Wigner function $f_{j}^{0}({\bf v})$ in a zero equilibrium electromagnetic field, one can set
\begin{eqnarray}
\label{e26}
f_j &=& f_{j}^{0}({\bf v}) + f_{j}^{1}({\bf v})\,\exp[i({\bf k}\cdot{\bf r}-\omega\,t)] \,,\\
\label{e24}
{\bf E} &=& {\bf E}_{1}\,\exp[i({\bf k}\cdot{\bf r}-\omega\,t)] \,,\\
\label{e25}
{\bf B} &=& {\bf B}_{1}\,\exp[i({\bf k}\cdot{\bf r}-\omega\,t)] \,,
\end{eqnarray}
where $f_{j}^{1}({\bf v}), {\bf E}_{1}$ and ${\bf B}_{1}$ are first order quantities. In this case it follows from Eqs. (\ref{e24}--\ref{e25}) that
\begin{equation}
\label{e28}
\tilde{\bf E}_j = {\bf E}\,L_j \,,\quad \tilde{\bf B}_j = {\bf B}\,L_j \,,
\end{equation}
defined in terms of the operators
\begin{equation}
\label{e29}
L_j = \frac{\sinh\theta_j}{\theta_j} \,, \quad \theta_j = \frac{\hbar}{2\,m_{j}}{\bf k}\cdot\frac{\partial}{\partial\,{\bf v}} \,.
\end{equation}
The operator $L_j$ in expression (\ref{e29}) is understood in a Taylor-expanded sense. For instance up to second-order in $\hbar$ we have
\begin{equation}
L_j = 1 + \frac{\hbar^2}{24 m_{j}^2}({\bf k}\cdot\frac{\partial}{\partial{\bf v}})^2 + \dots \,.
\end{equation}

Linearizing Eq. (\ref{e11}) we get
\begin{equation}
\label{f1}
f_{j}^1 = - \frac{i q_j}{m_j (\omega - {\bf k}\cdot{\bf v})}\,\left(\tilde{\bf E}_j + {\bf v}\times\tilde{\bf B}_j\right)\cdot\frac{\partial f_{j}^0}{\partial{\bf v}} \,,
\end{equation}
while the linearized Maxwell-Faraday and Amp\`ere-Maxwell equations resp. gives
\begin{eqnarray}
\label{e1}
{\bf k}\times{\bf E}_1 &=& \omega {\bf B}_1 \,,\\
\label{b1}
{\bf k}\times{\bf B}_1 &=& - i\mu_0\sum_j q_j n_{0j}\int d{\bf v} {\bf v} f_{j}^1 - \frac{\omega}{c^2} {\bf E}_1 \,,
\end{eqnarray}
where $\mu_0$ is the vacuum permeability constant, $c$ is the speed of light and $n_{0j}$ is the equilibrium number density of specie $j$. The normalization $\int\,d{\bf v}\,f_{j}^{0} = 1$ is applied. The line of the calculation is then very similar to the classical case \cite{BretReview2010}. Eliminating $f_{j}^1$ and ${\bf B}_1$ between Eqs. (\ref{f1})--(\ref{b1}) and using Eq. (\ref{e28}) yields ${\bf T}({\bf k},\omega)\cdot{\bf E}_1 = 0$, with
\begin{eqnarray}
T_{\alpha\beta} &=& \left(\frac{\omega^2}{c^2} - k^2\right)\delta_{\alpha\beta} + k_\alpha k_\beta \nonumber \\ \label{tab} &+& \sum_j \frac{\omega_{pj}^2}{c^2} \int d{\bf v} v_\alpha \left[L_j \frac{\partial f_{j}^0}{\partial v_\beta} + \frac{v_\beta}{\omega - {\bf k}\cdot{\bf v}} L_j \left({\bf k}\cdot\frac{\partial f_{j}^0}{\partial{\bf v}}\right)\right] \,,
\end{eqnarray}
where $\omega_{pj} = (n_{0j}q_{j}^2/(m_j \varepsilon_0))^{1/2}$ is the plasma frequency for specie $j$.

The expression (\ref{tab}) can be simplified using the two following properties,
\begin{eqnarray}
\label{e30}
L_j\left({\bf k}\cdot\frac{\partial f_{j}^{0}}{\partial\,{\bf v}}\right) &=& \frac{m_j}{\hbar}\,\left(f_{j}^{0}\left[{\bf v} + \frac{\hbar\,{\bf k}}{2\,m_j}\right] - f_{j}^{0}\left[{\bf v} - \frac{\hbar\,{\bf k}}{2\,m_j}\right]\right) \,, \\
\int d{\bf v} v_\alpha L_j \frac{\partial f_{j}^0}{\partial v_\beta} &=& - \delta_{\alpha\beta} \,,
\end{eqnarray}
which can be proven by series expansion of $L_j$ besides an integration by parts. In this way we find
\begin{equation}
\label{t}
{\bf T}({\bf k},\omega) = \frac{\omega^2}{c^2}\varepsilon({\bf k},\omega) + {\bf k}\otimes{\bf k} - k^2\,{\bf I} \,,
\end{equation}
where
\begin{eqnarray}
\label{eps}
\varepsilon_{\alpha\beta} &=& \delta_{\alpha\beta}\,\left(1 - \sum_{j}\,\frac{\omega_{pj}^2}{\omega^2}\right)  \\ \nonumber  &+& \int\,d{\bf v}\,\frac{v_\alpha\,v_\beta}{\hbar\,(\omega - {\bf k}\cdot{\bf v})}\,\sum_{j}\,\frac{m_{j}\,\omega_{pj}^2}{\omega^2}\,\left(f_{j}^{0}\left[{\bf v} + \frac{\hbar\,{\bf k}}{2\,m_j}\right] - f_{j}^{0}\left[{\bf v} - \frac{\hbar\,{\bf k}}{2\,m_j}\right]\right).
\end{eqnarray}

From Eq. (\ref{t}) the general dispersion relation for linear electromagnetic waves in quantum plasmas can be written as ${\rm det}(T_{\alpha\beta}) = 0$. As is well known, linear dispersion relations for quantum plasmas have been discussed for decades. For instance, Lindhard \cite{Lindhard} obtained both the longitudinal and the transverse dielectric tensors for a quantum electron gas in a Fermi-Dirac equilibrium. For arbitrary equilibria, the transverse and longitudinal dispersion relations for quantum plasmas have been found by Klimontovich and Silin \cite{Klimontovich} and Bohm and Pines \cite{BohmPines} respectively. In addition Silin and Rukhadze \cite{Silin} and Kuzelev and Rukhadze \cite{Kuzelev} have found the general dielectric tensor in Eq. (\ref{eps}) in a slightly different presentation, using the gauge-variant Wigner-Maxwell and Schr\"odinger-Maxwell systems respectively. Also Kelly \cite{Kelly} has found the dispersive properties of a quantum plasma under an homogeneous magnetic field, using a gauge-dependent Wigner formalism.
Furthermore, the straightforward derivation based on the Stratonovich-Serimaa gauge invariant Wigner-Maxwell system should be compared to the more cumbersome calculations needed in the gauge-dependent formalism \cite{Kelly}, applicable only in the particular case of homogeneous magnetic fields. In addition, Eqs. (\ref{t}, \ref{eps}) are valid for multi-species quantum plasmas and arbitrary equilibria, encompassing both longitudinal and transverse perturbations in an unified way.

\begin{figure}
\includegraphics[width=0.45\textwidth]{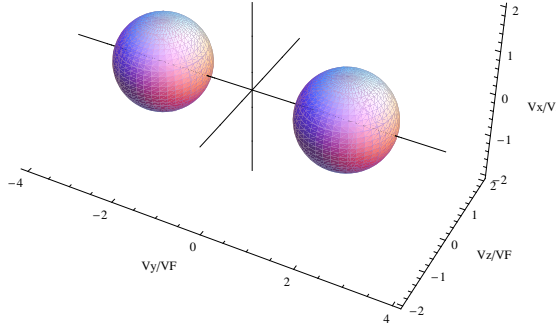}
\caption{(Color Online) Distribution function considered for two identical counter streaming beams with here $V_b=2V_F$.}
\label{fig1}
\end{figure}

\section{The filamentation instability}
\subsection{Dispersion Equation}

We now turn to investigate the filamentation instability with wave vector ${\bf k} = (k,0,0)$, for a beam-plasma system where the flow is aligned with the $y$ axis. For two symmetric identical counter streaming beams at $T=0$ with densities $n_b/2$ and velocity $V_b$, the dispersion equation reads \cite{BretPRE2004}
\begin{equation}\label{eq:disper}
\varepsilon_{yy} = k^2\,c^2/\omega^2.
\end{equation}
As can be seen in Figure \ref{fig1}, the distribution function $F$ is the sum of two spheres of radius $V_F$ (Fermi velocity) centered around $\pm V_b \mathbf{y}$,
\begin{equation}
f_{\pm}^{0}({\bf v}) = \frac{3}{4 \pi V_{F}^3} \quad {\rm if}\quad v_x^2 + (v_y \mp V_b)^2 + v_z^2 < V_{F}^2 \,,
\end{equation}
and  $f_{\pm}^{0} = 0$ outside the Fermi sphere. For this equilibrium a substitution of variables readily simplifies $\varepsilon_{yy}$ given by Eq. (\ref{eps}) into,
\begin{equation}
\varepsilon_{yy} = 1 - \frac{\omega_p^2}{\omega^2} - \frac{3 \omega_p^2 k^2}{4\pi\,V_F^3 \omega^2}\,\int_{v<V_F}\,d{\bf v}\,\frac{(v_y^2 + V_b^2)}{(\omega - k\,v_x)^2 - \hbar^2 k^4/(4 m^2)}.
\end{equation}
This quadrature can be calculated switching to spherical coordinate with
\begin{eqnarray}
v_x &=& v \cos \theta,~~(v,\theta)\in[0,V_F]\times[0,\pi] \nonumber\\
v_y &=& v \cos\varphi \sin\theta,~~\varphi\in [0,2\pi] \nonumber\\
v_z &=& v \sin\varphi \sin\theta.
\end{eqnarray}
The calculation is more easily performed integrating first over the variable $\varphi$, then over $\theta$ and finally over $v$. The final result is
\begin{equation}
\varepsilon_{yy} = 1 - \frac{\omega^2_p}{\omega^2}-\frac{\omega^2_p/\omega^2}{2^8\hbar k^5m^3V_F^3}\left[\mathcal{A}+\mathcal{B}\ln\left|\frac{\hbar k^2-2kmV_F-2m\omega}{\hbar k^2+2kmV_F-2m\omega}\right|
+\mathcal{C}\ln\left|\frac{\hbar k^2-2kmV_F+2m\omega}{\hbar k^2+2kmV_F+2m\omega}\right|\right],
\end{equation}
with,
\begin{eqnarray}
\mathcal{A} &=& 8 \hbar k^3 m V_F \left[3 \hbar^2 k^4-4 m^2 \left(k^2 \left(12 V_b^2+5 V_F^2\right)-9 \omega^2\right)\right],\\
\mathcal{B} &=& 3 \left[(\hbar k^2-2 m \omega)^2-(2 k m V_F)^2\right] \left[\hbar^2 k^4-4 \hbar k^2 m \omega+4 m^2 \left(\omega^2-k^2 \left(4 V_b^2+V_F^2\right)\right)\right],  \nonumber\\
\mathcal{C}&=& 3 \left[(\hbar k^2+2 m \omega)^2-(2 k m V_F)^2\right] \left[\hbar^2 k^4+4 \hbar k^2 m \omega+4 m^2 \left(\omega^2-k^2 \left(4 V_b^2+V_F^2\right)\right)\right].\nonumber
\end{eqnarray}

The classical cold plasma limit is correctly recovered setting $\hbar \rightarrow 0, V_F \rightarrow 0$,
\begin{equation}
\varepsilon_{yy} = 1 - \frac{\omega_p^2}{\omega^2} - \frac{\omega_p^2\,k^2\,V_b^2}{\omega^4} ,
\end{equation}
yielding with $\omega = i\delta$ the exact expression for the growth rate $\delta$,
\begin{equation}\label{eq:cold}
\delta^2 = \frac{1}{2} \left[c^2 k^2+\omega_p^2-\sqrt{4 k^2 V_b^2 \omega_p^2+\left(c^2 k^2+\omega_p^2\right)^2}\right]
\left\{ \begin{array}{ll}
=kV_b, & k\ll \omega_p/c,\\
=V_b\omega_p/c, & k\gg \omega_p/c,
\end{array} \right.
\end{equation}
in agreement with the results from the classical cold plasma model \cite{BretReview2010}.

The present calculation is worth comparing to the fluid theory including a Bohm pressure term \cite{BretPoPQuantum2007,BretPoPQuantum2008} and with $V_F \rightarrow 0$. In this fluid limit, the dispersion relation can be exactly solved and gives
\begin{equation}\label{eq:fluid}
\omega^2 = \frac{1}{2}\left[\omega_p^2 + c^2\,k^2 + \frac{\hbar^2\,k^4}{4\,m^2} - \left([\omega_p^2 + c^2\,k^2 - \frac{\hbar^2\,k^4}{4\,m^2}]^2 + 4\,k^2\,V_b^2\,\omega_p^2\right)^{1/2}\right]
\end{equation}
for the unstable mode. Knowing that the fluid and the kinetic models should merge in the long wave length limit, we first expand the above equation for small $k$,
\begin{equation}
\label{fluid}
\omega^2 = - k^2\,V_b^2  + \frac{\hbar^2\,k^4}{4\,m^2} + V_b^2\,(V_b^2 + c^2)\,\frac{k^4}{\omega_p^2} + \mathcal{O}(k^6) \,.
\end{equation}

To compare with the kinetic theory, we now expand $\varepsilon_{yy}$ in the same limit and with $V_F \rightarrow 0$,
\begin{equation}
\label{longw}
\varepsilon_{yy} = 1 - \frac{\omega_p^2}{\omega^2} - \frac{\omega_p^2\,k^2\,V_b^2}{\omega^4} - \frac{\hbar^2\,\omega_p^2\,k^6\,V_b^2}{4\,m^2\,\omega^6} + \mathcal{O}(k^8) \,.
\end{equation}
The dispersion equation (\ref{longw}) is a cubic  for $\omega^2$. There are analytic formulas which can be applied, but in the long wavelength limit it is cheaper to solve it recursively. The result agrees with Eq. (\ref{fluid}), confirming the equivalence between quantum kinetic and quantum fluid models for long wavelengths. Such equivalence can be checked on Figure \ref{fig3} which compares the kinetic with the quantum fluid with Bohm potential and the classical cold plasma calculations.

\begin{figure}
\includegraphics[width=0.45\textwidth]{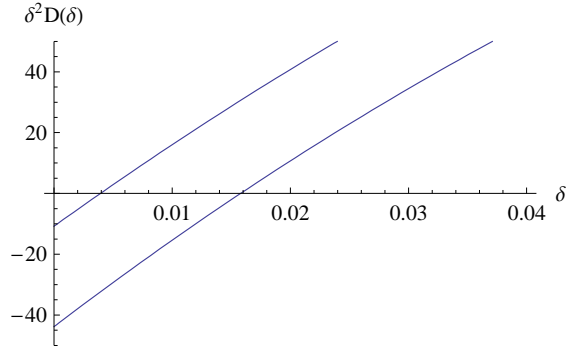}
\caption{Typical plot of $\delta^2D(\delta)$, where $D$ is the dispersion function and $\Omega=0+i\delta$, for $Z=1.6$ (lower curve) and $Z=1.7$ (upper curve). The intersections with the horizontal axis correspond to the growth rate. Parameters are $\beta=0.1$ and $\rho=10$.}
\label{fig2}
\end{figure}

\begin{figure}
(a)\\
\includegraphics[width=0.45\textwidth]{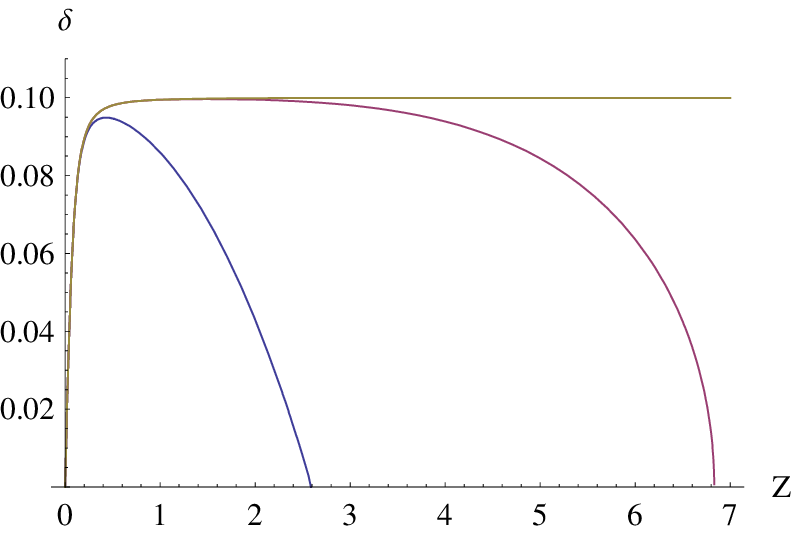}\\
(b)\\
\includegraphics[width=0.45\textwidth]{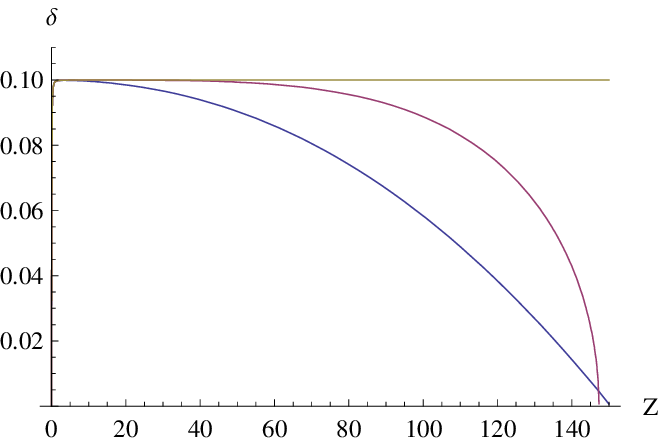}\\
\caption{(Color Online) Kinetic growth rate (blue) obtained solving the kinetic dispersion equation, compared to the quantum fluid result with Bohm pressure term (purple) and to the classical cold plasma result (yellow), in terms of the reduced wave vector $Z$. In (a), parameters are $\beta=0.1$ and $\rho=15$, and the fluid unstable range is wider than the kinetic one. In (b), parameters are $\beta=0.1$ and $\rho=900$, and the fluid unstable range is \emph{smaller} than the kinetic one. The saturation value for the classical cold curve is simply $\beta$.}
\label{fig3}
\end{figure}

In order to proceed further in the calculations, we introduce the usual dimensionless parameters \cite{Haas2000},
\begin{equation}
\label{dim}
    \Omega=\frac{\omega}{\omega_p},\,\,Z=\frac{k V_b}{\omega_p},\,\,
    \beta=\frac{V_b}{c},\,\,H=\frac{\hbar \omega_p}{mV_b^2},\,\, \rho=\frac{V_b}{V_F}.
\end{equation}
In the classical case, accounting for a thermal spread requires an additional parameter, namely, the temperature. Here, the thermal spread $V_F$ is related to the density. As a result, the above parameters are not independent of each others and one can check that
\begin{equation}\label{eq:H}
    H=\frac{2\sqrt{\alpha_c}}{\sqrt{3\pi\beta}\rho^{3/2}},
\end{equation}
where $\alpha_c=e^2/\hbar c=1/137$ is the fine-structure constant. Like in the classical case \cite{BretReview2010},  the roots of the dispersion equation are here found with zero real part. Setting $\Omega=0+i\delta$ and denoting $D(\delta)$ the dispersion function yielding the dispersion equation $D(\delta)=0$, Figure \ref{fig2} sketches a typical plot of $\delta^2D(\delta)$ for two values of $Z$. The intersection of the curves with the horizontal axis directly gives the growth rate.

Figure \ref{fig3} compares the kinetic growth rate obtained solving numerically the full dispersion equation with the classical cold and the quantum fluid with Bohm term results for the parameters specified in caption. The kinetic equation has been solved using \emph{Mathematica}'s ``FindRoot'' routine, giving the fluid growth rate (\ref{eq:fluid}) as an initial guess. The agreement in the long wave length limit is clear. The classical cold plasma result saturates at the value $\beta$ for large $Z$ while the quantum fluid and the present kinetic results exhibit cutoffs, as the kinetic pressure eventually acts to prevent the pinching of small filaments.

Note on Fig. \ref{fig3}b that the range of unstable modes is \emph{smaller} in the fluid than in the kinetic case. We thus turn now to the investigation of the cutoff wave vector.

\subsection{Cutoff wave vector}
The cutoff wave number $Z_m$ can be found writing $\delta^2D(\delta)=0$ for $Z=Z_m$ and $\delta=0$. Denoting $L=\lim_{\delta=0}\delta^2D(\delta)$, one finds (only the numerator is shown)

\begin{eqnarray}\label{eq:L}
L=&&4 H Z \rho  \left[12 \beta ^2 (1-4 \rho ^2)+Z^2 (32+3 H^2 \beta ^2 \rho ^2)\right]\\
&+&3 \beta ^2 \left[16-8 (H^2 Z^2-8) \rho ^2+H^2 Z^2 (H^2 Z^2-16) \rho ^4\right] \ln\left|\frac{2-H Z \rho }{2+H Z \rho }\right|.\nonumber
\end{eqnarray}

We start searching the expression of the largest unstable wave vector $Z_m$ in the most interesting limit. This is the kinetic one, with  $\rho=V_b/V_F\rightarrow 0$, as the opposite limit is just the fluid one. By developing the logarithm in Eq. (\ref{eq:L}),  the equation $L=0$ simplifies  to
\begin{equation}\label{eq:ZmKin}
0=384 \beta ^2 \rho ^2+Z^2 \left[3 H^2 \beta ^2 \rho ^2 \left((H^2 Z^2-16) \rho ^2-12\right)-128\right].
\end{equation}
Assuming $H^2 Z^2\ll 16$ (to be checked later), we find directly
\begin{equation}\label{eq:ZmKinOK}
Z_m=\frac{2 \sqrt{6 \pi }\beta \rho^{3/2}}{\sqrt{8\pi\rho+\alpha_c\beta(3+4\rho^2)}}.
\end{equation}
We can now check our assumption $H^2 Z^2\ll 16$ . Replacing $Z$ by the value of $Z_m$ above, we find
\begin{equation}
H^2 Z_m^2=\frac{32\alpha_c\beta}{8\pi\rho +\alpha_c\beta(3+4\rho^2)}<32/3.
\end{equation}
It thus turns out that the condition $H^2 Z_m^2\ll 16$ is only weakly verified. However, Eq. (\ref{eq:ZmKin}) can be solved exactly very easily, and the solution found is numerically very close to Eq. (\ref{eq:ZmKinOK}).

Examining now the condition to expand the logarithm $HZ_m\rho\ll 2$, we can find the validity domain of Eq. (\ref{eq:ZmKinOK}),
\begin{equation}\label{eq:CondRho}
H Z_m \rho=\rho\sqrt{\frac{32\alpha_c\beta}{8\pi\rho+\alpha_c\beta(3+4\rho^2)}}\ll 2 \Leftrightarrow \rho \ll \frac{2\pi+\sqrt{4\pi^2+3\alpha_c\beta^2}}{2\alpha_c\beta}\sim\frac{2\pi}{\alpha_c\beta}\sim\frac{860}{\beta}.
\end{equation}
Equation (\ref{eq:ZmKinOK}) is thus found valid in a very wide range of parameter defined by the strong inequality (\ref{eq:CondRho}). Note worthily, it defines various $\rho$ scalings. For $\rho\ll\alpha_c\beta/8\pi$, $Z_m\propto\rho^{3/2}$. Then, the denominator behaves like $\sqrt{\rho}$, yielding a $Z_m\propto\rho$ scaling until the quadratic term under the square root overcomes the linear one. Comparing these two terms gives a criterion on $\rho$ almost identical to Eq. (\ref{eq:CondRho}).

For $\rho\gg 2\pi/\alpha_c\beta$, the logarithm in Eq. (\ref{eq:L}) can be expanded assuming now $HZ_m\rho\gg 1$. Performing such expansion and replacing $H$ by its value given by Eq. (\ref{eq:H}) gives the following equation for $Z_m$,

\begin{equation}
8 Z_m^4 \alpha_c+9 Z_m^2 \alpha_c \beta ^2-9 \pi  \beta ^3 \rho (1+4 \rho ^2)=0.
\end{equation}

This equation can be solved exactly. Expanding the relevant root for large $\rho$ gives,

\begin{equation}\label{eq:ZmKinLarge}
Z_m=\left(\frac{9 \pi  }{2 \alpha_c}\right)^{1/4} (\beta\rho)^{3/4}.
\end{equation}

Equations (\ref{eq:ZmKinOK},\ref{eq:ZmKinLarge}) eventually define three different scalings which are summarized in Table \ref{tab:summary}. It is interesting to ``unfold'' the dimensionless parameters in order to explain the key quantity $HZ\rho$. We find,
\begin{equation}
HZ\rho=\frac{\hbar\omega_p}{mV_b^2}\frac{kV_b}{\omega_p}\frac{V_b}{V_F}=\frac{\hbar k}{mV_F}=\frac{k}{k_F},
\end{equation}
where $k_F$ is the Fermi wave number.

\begin{table}[h]
\caption{\label{tab:summary}Analytical expressions for the largest unstable wave vector in the various regimes.}
\begin{ruledtabular}
\begin{tabular}{lccc}
$\rho$  & $0<\rho<\frac{\alpha_c\beta}{8\pi}$   & $\frac{\alpha_c\beta}{8\pi}<\rho<\frac{2\pi}{\alpha_c\beta}$  & $\frac{2\pi}{\alpha_c\beta}<\rho$ \\
\hline
$Z_m$  &   $\sqrt{\frac{8 \pi \beta  }{\alpha_c}}\rho^{3/2}$                &   $\sqrt{3}\beta\rho$                                     &    $\left(\frac{9 \pi  }{2 \alpha_c}\right)^{1/4} (\beta\rho)^{3/4}$   \\
\end{tabular}
\end{ruledtabular}
\end{table}

\begin{figure}
\includegraphics[width=0.45\textwidth]{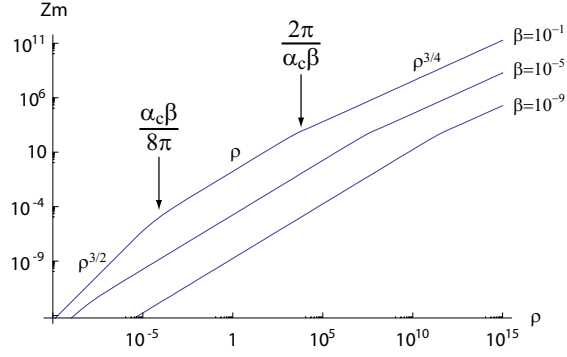}
\caption{(Color Online) Most unstable wave vector $Z_m$ in terms of $\rho$. The analytical expressions in the various regimes are provided on Table \ref{tab:summary}.}
\label{fig4}
\end{figure}

Figure \ref{fig4} displays the numerical evaluation of the cutoff wave number in terms of $\rho$ and for 3 values of $\beta$. The analytical expression reported in Table \ref{tab:summary} cannot be distinguished from the numerical calculation within their range of validity.

Let us finally compare the kinetic cutoff with the fluid one. The largest unstable mode can be expressed exactly in the fluid model with the Bohm pressure term as \cite{BretPoPQuantum2007}, \begin{equation}
Z_{mf}=\frac{\beta}{\sqrt{2}}\left[\sqrt{1+8/H^2\beta^2}-1\right]^{1/2}.
\end{equation}
Replacing $H$ by its value in terms of $\rho$ from Eq. (\ref{eq:H}) and expanding for small and large $\rho$'s yield,

\begin{equation}\label{eq:cold_cut}
Z_{mf} =
\left\{ \begin{array}{ll}
\sqrt{\frac{3\pi\beta}{2\alpha_c}}\rho^{3/2}, & \rho\ll \left(\frac{\alpha_c\beta}{6\pi}\right)^{1/3},\\
\left(\frac{3\pi}{2\alpha_c}\right)^{1/4}(\beta\rho)^{3/4}, & \rho\gg \left(\frac{\alpha_c\beta}{6\pi}\right)^{1/3}.
\end{array} \right.
\end{equation}

These extreme scalings are thus very similar to the kinetic ones explained on Table \ref{tab:summary} in the limits $\rho\rightarrow 0,\infty$. The kinetic cutoff displays three different regimes and the fluid one above, only two. While in the intermediate kinetic regime, the kinetic cutoff is smaller than the fluid one (as in Fig. \ref{fig3}a), Table \ref{tab:summary} and Eq. (\ref{eq:cold_cut}) show that surprisingly, the fluid cutoff can be \emph{smaller} than the kinetic one (as in Fig. \ref{fig3}b).

\subsection{Most unstable wave vector and largest growth rate}
The most unstable wave vector $Z_{max}$ in the range $[0,Z_m]$, together with its growth rate $\delta_m$, are key quantities which eventually define the strength of the instability and its time scale. Although Fig. \ref{fig1} suggests otherwise, an analytical approach based on a Taylor expansion near $\Omega=0$ of the dispersion equation has not been found valid here. It seems that indeed, the expansion up to the second order is not enough to approach the numerical calculations.

\begin{figure}
\includegraphics[width=0.45\textwidth]{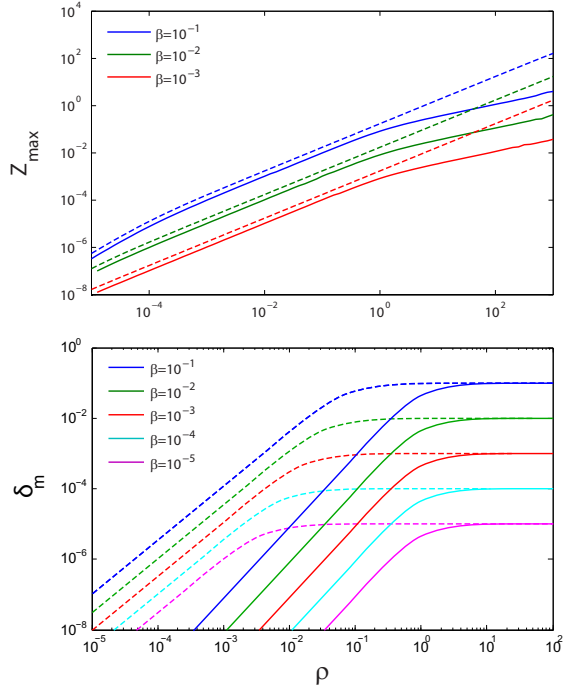}
\caption{(Color Online) Numerical determination of the most unstable wave vector $Z_{max}$ and its growth rate $\delta_m$, in terms of $\rho$ and for various values of $\beta$. Upper plot: Kinetic value of  $Z_{max}$ (plain curves) vs. Eq. (\ref{eq:ZmKinOK}) (dashed curves). Lower plot: Kinetic value of the largest growth rate $\delta_m$ (plain curves) compared to its fluid value (dashed curves).}
\label{fig5}
\end{figure}

We thus resort to a systematic exploration of the parameters phase space $(\beta,\rho)$ in order to extract scaling laws. The results are sketched on Figure \ref{fig5} within the parameter range where numerical stability allowed to derive trustful results.

The curves for the maximum growth rate all saturates at $\delta_m=\beta$ for $\rho\gg 1$. In the kinetic regime $\rho\ll 1$, simple scalings are evidenced in terms of the parameters, and the following fit has been found
\begin{equation}\label{eq:Dm}
  \delta_m \sim 0.72\beta\rho^2.
\end{equation}

The largest unstable wave vector $Z_m$, as given by Eq. (\ref{eq:ZmKinOK}), has been represented with the most unstable one $Z_{max}$ on Fig. \ref{fig5}. We obviously find fulfilled the inequality $Z_m<Z_{max}$. These two quantities remain locked to each other until they decouple from $\rho*(\beta)\sim 1$ slightly varying with $\beta$. For $\rho\ll \rho*$, the following equality is fulfilled with remarkable constancy, regardless of the value of $\beta$
\begin{equation}\label{eq:ZmSmall}
  Z_{max}(\rho<\rho*) \sim 0.6 \, Z_m.
\end{equation}

For $\rho\gg \rho*$, Fig. \ref{fig5} shows $Z_{max}$ switches from a simple $\rho$ to a measured $\rho^{1/2}$ scaling. As expected then, the fastest growing wave number tends to infinity, but slower than the cutoff. A typical plot in this range is displayed in Fig. \ref{fig3}a, where the growth rate quickly reaches its maximum $\sim\beta$ before its progressively comes down to zero.

Still on Fig. \ref{fig5}, the lower plot representing the largest growth rates shows the kinetic one undergoes a transition from $\delta_m\propto\rho^2$ to $\delta_m\sim\beta$ near $\rho=1$. The fluid results displays a measured $\rho^{3/2}$ until $\rho\sim\beta^{1/2}$, from where it saturates also at $\delta_m\sim\beta$. If then, one wishes to define the limit of the fluid model through the correspondence of the maximum fluid and kinetic growth rates, the fluid approximation is found valid for $\rho\gg 1$.

\section{Conclusion and discussion}
In conclusion, using Stratonovich's gauge invariant Wigner equation we have re-derived the general form of the electromagnetic dielectric tensor in a quantum plasma. The quantum filamentation instability was then treated using kinetic theory. The equilibrium Wigner function was taken as a pair of zero-temperature Fermi-Dirac distributions centered at the beams velocities. In this way, not only the quantum diffraction effects inherent to the kinetic equation (\ref{e11}) but also the fermionic character of the beams were included. The results were compared to the zero-temperature classical and quantum hydrodynamic equations (with $V_F \rightarrow 0$), showing agreement in the long-wavelength limit. In addition, analytical expressions are derived for the the largest growth rate, the cutoff wave vector and the most unstable wave vector.

It is worth to comment on the influence of the quantum properties against the filamentation instability. As apparent from Eq. (\ref{dim}), the quantum statistical effects decrease with the quantity $\rho = V_b/V_F$. Also quantum diffraction effects represented by the parameter $H = \hbar\omega_p/(m V_B^2)$ decrease with $\rho$, as follows from Eq. (\ref{eq:H}) for a fixed beam velocity. On the other hand, from Figs. (\ref{fig4}, \ref{fig5}) we see that the cuttof wavenumber, the largest unstable wave vector and its growth rate increases with $\rho$. Hence kinetic theory shows that quantum mechanics has a stabilizing r\^ole to the filamentation instability. The present theory can be applied to very dense counter-streaming plasmas obeying Fermi-Dirac statistics and where quantum diffraction can be  significant. 

\begin{acknowledgments}
This work was supported by projects ENE2009-09276 of the
Spanish Ministerio de Educaci\'{o}n y Ciencia and PEII11-0056-1890 of
the Consejer\'{i}a de Educaci\'{o}n y Ciencia de la Junta de
Comunidades de Castilla-La Mancha, and by CNPq (Conselho Nacional de Desenvolvimento Cient\'{\i}fico e Tecnol\'ogico).
\end{acknowledgments}


\begin{thebibliography}{99}
\bibitem{zutic} I. Zutic, J. Fabian and S. Das Sarma, Rev. Mod. Phys. {\bf 76}, 323 (2004). 
\bibitem{eliasson} B. Eliasson, P. K. Shukla and M. E. Dieckmann, New J. Phys. {\bf 8}, 55 (2006). 
\bibitem{uspekhi} P. K. Shukla and B. Eliasson, Phys. Uspekhi {\bf 53}, 51 (2010).
\bibitem{BretPoPQuantum2007} A. Bret, Phys. Plasmas {\bf 13}, 084503 (2007).
\bibitem{BretPoPQuantum2008} A. Bret, Phys. Plasmas {\bf 15}, 022109 (2008).
\bibitem{Stratonovich} R. L. Stratonovich, Sov. Phys.-Dokl. {\bf 1}, 414 (1956). 
\bibitem{Ichimaru} S. Ichimaru, {\it Basic Principles of Plasma Physics} (W. A. Benjamin, Inc., Reading, Massachusetts, 1973).
\bibitem{Clemmow} P. C. Clemmow and J. P. Dougherty, {\it Electrodynamics of Particles and Plasmas} (Reading, Mass.: Addison-Wesley Publ. Co., 1990).
\bibitem{Wigner} E. P. Wigner, Phys. Rev. {\bf 40}, 749 (1932). 
\bibitem{HaasNJP2010} F Haas, J Zamanian, M Marklund and G Brodin, New J. Phys. {\bf 12}, 073027 (2010).
\bibitem{Serimaa} O. T. Serimaa, J. Javanainen and S. Varr\'{o}, Phys. Rev. A {\bf 33}, 2913 (1986). 
\bibitem{BretReview2010} A. Bret, L. Gremillet and M. E. Dieckmann, Phys. Plasmas {\bf 17}, 120501 (2010).
\bibitem{Lindhard} J. Lindhard, K. Dan. Vidensk. Selsk. Mat. Fys. Medd. {\bf 28}, 1 (1954). 
\bibitem{Klimontovich} Y. L. Klimontovich and V. P. Silin, Zh. Eksp. Teor. Fiz. {\bf 23}, 151 (1952).
\bibitem{BohmPines} D. Bohm and D. Pines, Phys. Rev. {\bf 92}, 609 (1953). 
\bibitem{Silin} V. P. Silin and A. A. Rukhadze, {\it Elektromagnitnye Svoystva Plazmy i
Plazmopodobnykh Sred} (Moscow, Gosatomizdat, 1961).
\bibitem{Kuzelev} M. V. Kuzelev and A. A. Rukhadze, Phys. Uspekhi {\bf 42}, 603 (1999). 
\bibitem{Kelly} D. C. Kelly, Phys. Rev. {\bf 134}, A643 (1964). 
\bibitem{BretPRE2004} A. Bret, M.-C. Firpo and C. Deutsch, Phys. Rev. E {\bf 70}, 046401 (2004). 
\bibitem{Haas2000} F. Haas, G. Manfredi and M. Feix,
Phys. Rev. E {\bf 62}, 2763 (2000).
\end{thebibliography}
\end{document}